\documentclass[%
 reprint,
 amsmath,amssymb,
 aps,
]{revtex4-2}

\usepackage{graphicx}
\usepackage{dcolumn}
\usepackage{bm}
\usepackage{tikz}
\usetikzlibrary{positioning,arrows.meta}
\usepackage{xcolor}

\begin{document}

\preprint{APS/123-QED}

\title{Simulation-based forward modelling of the ACT thermal Sunyaev--Zel'dovich profile around DESI luminous red galaxies}

\author{Alicja Brzezowska}
\email{ab2927@cam.ac.uk}
\affiliation{%
 Cavendish Laboratory, JJ Thomson Avenue, CB3 0US, Cambridge, UK
}%
\author{Boryana Hadzhiyska}%
 \email{boryanah@ast.cam.ac.uk}
\affiliation{%
Institute of Astronomy, Madingley Road, Cambridge, CB3 0HA, UK
}%
\affiliation{%
Kavli Institute for Cosmology Cambridge, Madingley Road, Cambridge, CB3 0HA, UK
}%


\begin{abstract}
The thermal Sunyaev--Zel'dovich (tSZ) effect probes the thermodynamic state of diffuse ionized gas by tracing the integrated electron pressure around galaxies and clusters. Interpreting tSZ measurements in halo outskirts is, however, challenging, because the measured signal receives substantial contributions from correlated large-scale structure. This requires modelling both the halo gas profiles and their surrounding matter distribution. In this work, we develop the public code \texttt{AbacustSZ.py}, a simulation-based framework for generating tSZ maps from the light cones of the \textsc{Abacus} \textit{N}-body simulation. \texttt{AbacustSZ.py} paints generalized Navarro--Frenk--White (gNFW) pressure profiles onto light-cone halos, retaining the two-halo contribution directly from the underlying simulation. Furthermore, it can preserve the projected morphology of each halo, using elliptical footprints derived from the halo shapes and redistributing the signal according to the local dark-matter distribution. We apply \texttt{AbacustSZ.py} to the stacked ACT tSZ profiles around DESI photometric luminous red galaxies (LRGs). As \texttt{AbacustSZ.py} is optimised for fast evaluation of different pressure-profile scenarios, we perform 130 forward model evaluations, and a Gaussian-process emulator interpolates the predictions across the gNFW parameter space for Bayesian inference. We show that the data prefer a much shallower, smaller pressure profile than the fiducial calibration of Battaglia et al. (2012).

\end{abstract}

\maketitle


\section{Introduction\label{sec:intro}}

Baryons constitute approximately 16\% of the total matter content of the Universe \cite{Planck2018_VI}, yet only a small fraction of them is found in stars and galaxies \cite{Fukugita1998,2004ApJ...616..643F}.
Most are expected to reside in diffuse ionized gas, such as the warm--hot intergalactic medium (WHIM) \cite{Cen1999,Cen_2006}. Although the global baryon census has now largely been closed \cite{Shull2012,Macquart2020}, it remains poorly constrained where these baryons reside relative to dark matter halos, and how far feedback from galaxies and active galactic nuclei redistributes them.

One of the most powerful observational probes of diffuse ionized 
baryons is the Sunyaev--Zel'dovich (SZ) effect, which arises from the inverse Compton scattering of Cosmic Microwave Background (CMB) photons by free electrons along the line of sight \cite{BIRKINSHAW199997, Zeldovich1969}. It is commonly divided into two components: the thermal 
Sunyaev--Zel'dovich (tSZ) effect, which probes the integrated thermal 
pressure of ionized gas, and the kinetic Sunyaev--Zel'dovich (kSZ) 
effect, which probes the electron momentum field.

Recent kSZ and some X-ray studies have provided growing evidence that a substantial fraction of baryons reside beyond halo virial radii, favouring scenarios in which baryonic feedback redistributes gas far into the circumgalactic and intergalactic medium \cite{hadzhiyska2025evidencelargebaryonicfeedback, Bigwood2024, RiedGuachalla2025, Schaan_2021, 2025MNRAS.540..143M, Hadzhiyska2025lensing, 2026MNRAS.549ag993S, Qu2026, 2026PhRvD.114h3541H, 2026A&A...707A.362P}. However, constraining the thermal state of this redistributed gas additionally requires modelling the tSZ signal \cite{Popik2025, 2026arXiv260800164S}. Joint kSZ and tSZ analyses are powerful, because they provide complementary constraints on the density, temperature, and feedback energetics of diffuse baryons \cite{Schaan_2021, Amodeo_2021, Vavagiakis2021, Popik2025, 2026PhRvD.114j3513M}. Motivated by these developments, tSZ profiles have recently been measured around the same DESI Luminous Red Galaxy (LRG) samples as the kSZ analyses \cite{liu2025measurementsthermalsunyaevzeldovicheffect}. Interpreting them requires accurate modelling of the tSZ signal around these galaxies.

Accurate modelling of the tSZ signal is further motivated 
by the demands of high-precision cosmological surveys, such as the Vera C. Rubin Observatory and Euclid \cite{Laureijs2011Euclid,Ivezic2019LSST}. 
Baryonic feedback suppresses the matter power spectrum on nonlinear scales and can bias inferred cosmological parameters such as 
$\sigma_8$ and $\Omega_m$, if unaccounted for \cite{Schneider_2020,Pandey2023,McCarthy2023FLAMINGO,Bigwood2024, Amon2022}. 

Modelling the tSZ signal around LRGs is however non-trivial. The stacked signal receives contributions not only from gas gravitationally bound to the target halos, but also from correlated neighbouring large-scale structure, the so-called two-halo term, which increasingly dominates the observed signal at large projected separations. Cosmological hydrodynamical simulations, whilst most physically complete, are limited by their computational cost \cite{McCarthy_2016,Nelson2021IllustrisTNG,Schaye2023FLAMINGO}. At present, large suites of feedback variations are mostly available in small volumes \cite{Villaescusa_Navarro_2021}, whereas the Gpc-scale volumes needed to sample the massive halos which dominate the tSZ signal exist for discrete feedback variants \cite{Schaye2023FLAMINGO,Kugel2023FLAMINGO}. Fast semi-analytical halo-based approaches, by contrast, rely on simplified analytical treatments of the one-halo to two-halo transition, which becomes insufficient for studies dedicated to halo outskirts \cite{Amodeo_2021,Moser_2021,Hill_2018}. An intermediate class of approaches has emerged, augmenting large \textit{N}-body simulations with baryonic prescriptions \cite{Osato_2022,schneider2025baryonificationalternativehydrodynamicalsimulations,Kovac2025}. These are far cheaper than hydrodynamical simulations, and retain the two-halo contribution of the underlying large-scale structure. 

These approaches can be broadly divided into two classes. In baryonification, the particles of each halo are displaced such that its matter profile follows a parametrized model of dark matter, stars and gas, calibrated on hydrodynamical simulations \cite{SchneiderTeyssier2015,Schneider2019,Arico2020}. Originally developed to model the baryonic suppression of the matter power spectrum, baryonification has since been extended to the gas pressure and the tSZ effect \cite{Arico2024tSZ,Anbajagane2024,schneider2025baryonificationalternativehydrodynamicalsimulations,Kovac2026}, and has been fitted to weak-lensing, X-ray and kSZ data \cite{Schneider2022,Arico2023,Kovac2025}. Painting, or pasting, methods instead leave the matter field unchanged, and place a gas or pressure profile at the position of each halo. This profile can either be analytic, as in full-sky simulations of the microwave sky \cite{Sehgal2010,Stein2020Websky,Omori2024Agora,2025ascl.soft03033L}, derived from gas in hydrostatic equilibrium within the halo potential \cite{Osato_2022,Keruzore2024,Lau2025Uchuu}, or learned from hydrodynamical simulations \cite{Troster2019,Thiele2020}. However, these tools have mostly been used to generate mock tSZ observables at fixed, calibrated parameters, rather than to fit the pressure profile to tSZ measurements.

In this work, we develop \texttt{AbacustSZ.py}, a simulation-based framework for generating mock tSZ maps from the light-cone outputs of an \textsc{Abacus} \textit{N}-body simulation. Analytic gas-pressure profiles based on the generalised Navarro--Frenk--White (gNFW) form \cite{Nagai2007} are painted at the positions of light-cone halos, and the particle maps are used to redistribute the signal within each halo, following the underlying matter distribution. The code is optimized for parallel painting of multiple gNFW parameter prescriptions. We apply the framework to forward model recent stacked tSZ measurements around DESI LRGs \cite{liu2025measurementsthermalsunyaevzeldovicheffect}, and obtain constraints on the gas pressure profile.

Section~\ref{sec:theory} summarizes the theoretical background of the tSZ effect and the analytic pressure model employed in this work. Section~\ref{sec:abacustsz} presents the \texttt{AbacustSZ.py} framework and its features. Section~\ref{sec:modelling} applies it to the stacked ACT DR5 $\times$ DESI LRG measurement, covering the data, the forward model and the resulting constraints on the pressure profile. Conclusions are raised in section~\ref{sec:conclusions}. 

\section{Theory\label{sec:theory}}

\subsection{Thermal Sunyaev-Zel'dovich effect}

The tSZ effect is a secondary anisotropy of the CMB, caused by inverse Compton scattering of CMB photons off hot free electrons, which results in a frequency-dependent distortion of the blackbody spectrum \cite{Sunyaev1972,BIRKINSHAW199997,Carlstrom2002}. Physically, electrons moving with different thermal velocities Doppler-boost the scattered photons, and averaging over the thermal velocity distribution of the electrons leads to a characteristic $y$-type spectral distortion, whose amplitude is proportional to the electron temperature. In the non-relativistic limit, the resulting temperature change is
\begin{equation}
    \frac{\Delta T_{\rm tSZ}(\hat{\boldsymbol{n}})}{T_{\text{CMB}}} = y(\hat{\boldsymbol{n}})\, f_{\rm tSZ}(\nu),
\end{equation}
where $f_{\rm tSZ}(\nu) = x \coth(x/2) - 4$, $x \equiv h\nu/(k_B T_{\mathrm{CMB}})$ is evaluated at the effective frequency of the band, and the Compton-$y$ parameter $y(\hat{\boldsymbol{n}})$ is the integral of the electron pressure along the line of sight:

\begin{equation}
\label{eq:compton_y}
\begin{aligned}
y(\hat{\boldsymbol{n}})
&=
\frac{\sigma_T k_B}{m_e c^2}
\int dl \,
n_e(\mathbf{r}) T_e(\mathbf{r}) \\
&=
\frac{\sigma_T}{m_e c^2}
\int \frac{d\chi}{1+z}
P_e(\chi \hat{\boldsymbol{n}}, z),
\end{aligned}
\end{equation}

where $\sigma_T$ is the Thomson cross-section, $m_e$ the electron mass, $n_e$, $T_e$ and $P_e=n_e k_B T_e$ the (physical) electron density, temperature and pressure, $\chi$ the comoving distance, and $dl=d\chi/(1+z)$ the physical line-of-sight element.

Under the assumption of self-similar gravitational collapse, halos form as approximately scaled versions of one another, leading to the characteristic scaling of the integrated signal, $Y D_A^2 \propto E(z)^{2/3} M^{5/3}$, where $Y=\int y\,d\Omega$, $D_A$ is the angular-diameter distance and $E(z)=H(z)/H_0$ \cite{Kaiser1986}. The tSZ signal is therefore strongly weighted towards the most massive halos in the sample.


\subsection{Pressure Profile}

Hydrodynamical simulations and X-ray observations have shown that halo pressure profiles are well described by generalized Navarro--Frenk--White (gNFW) profiles \cite{Nagai2007,Arnaud_2010,Battaglia_2012}. These extend the NFW description of dark matter halos \cite{Navarro1997UniversalProfile} by introducing additional flexibility in the profile shape, allowing them to capture the impact of baryonic processes such as radiative cooling and feedback. In this work, we adopt the parameterization of Battaglia et al.~\cite{Battaglia_2012}:
\begin{equation}
\label{eq:gnfw}
\frac{P_{\rm th}(r)}{P_{200c}}
=
P_0
\left(\frac{x}{x_c}\right)^{\gamma}
\left[
1+\left(\frac{x}{x_c}\right)^{\alpha}
\right]^{-\beta},
\end{equation}
where $x\equiv r/R_{200c}$, $P_0$ controls the overall pressure normalization and $x_c$ sets the characteristic scale radius. The profile has a logarithmic slope of $\gamma$ at $x\ll x_c$ and of $\gamma-\alpha\beta$ at $x\gg x_c$.
The parameter $\alpha$ sets the sharpness of the transition between the two regimes \cite{Nagai2007}. The profile is expressed in units of the self-similar pressure
\begin{equation}
P_{200c} = \frac{G M_{200c}\, 200\, \rho_{\rm crit}(z)\, f_b}{2 R_{200c}},
\end{equation}
where $\rho_{\rm crit}(z)=3H^2(z)/(8\pi G)$ is the critical density, $M_{200c}$ and $R_{200c}$ denote the mass and radius enclosing a mean density of $200\rho_{\rm crit}(z)$, and $f_b=\Omega_b/\Omega_m$ is the cosmic baryon fraction. The electron pressure entering Eq.~(\ref{eq:compton_y}) is $P_e = [(2X+2)/(5X+3)]\,P_{\rm th} \simeq 0.518\,P_{\rm th}$ for a fully ionized gas with primordial hydrogen mass fraction $X=0.76$ \cite{Battaglia_2012}.

Furthermore, hydrodynamical simulations show that the gNFW parameters retain a residual mass and redshift dependence, reflecting departures from purely self-similar evolution \cite{Battaglia_2012}. Battaglia et al.~\cite{Battaglia_2012} parameterize this behaviour for $P_0$, $x_c$ and $\beta$ using power-law scalings:
\begin{equation}
\label{eq:gnfw_scaling}
A(M_{200c}, z)
=
A_0
\left(\frac{M_{200c}}{10^{14} \, M_{\odot}}\right)^{\alpha_m}
(1+z)^{\alpha_z},
\end{equation}
where $A_0$, $\alpha_m$, and $\alpha_z$ describe the normalization, mass dependence, and redshift evolution of each parameter, whilst $\alpha$ and $\gamma$ were held fixed. Table~\ref{tab:battaglia_parameters} lists the fiducial values used for the mock maps of Sec.~\ref{sec:abacustsz}, which correspond to the AGN-feedback calibration of Battaglia et al.~\cite{Battaglia_2012}. When fitting the data, we vary the amplitudes $P_0$ and $\beta_0$ together with $\alpha$ and $\gamma$ (Sec.~\ref{sec:forward_model}).




\begin{table}[h]
\centering
\begin{tabular}{cccc}
\hline
\textbf{Parameter} & $A_0$ & $\alpha_m$ & $\alpha_z$ \\
\hline
$P_0$    & 18.1  & 0.154    & $-0.758$ \\
$x_c$    & 0.497 & $-0.00865$ & 0.731 \\
$\beta$  & 4.35  & 0.0393   & 0.415 \\
$\alpha$ & 1.0   & --       & -- \\
$\gamma$ & $-0.3$ & --      & -- \\
\hline
\end{tabular}
\caption{
Fiducial gNFW pressure-profile parameters of Battaglia et al.~\cite{Battaglia_2012}, with the mass and redshift scalings of Eq.~(\ref{eq:gnfw_scaling}).
}
\label{tab:battaglia_parameters}
\end{table}

\section{\texttt{AbacustSZ.py}\label{sec:abacustsz}}

\begin{figure*}[t]
\centering

\resizebox{\textwidth}{!}{%
\begin{tikzpicture}[
    >=Stealth,
    line width=0.8pt,
    box/.style={
        draw,
        rectangle,
        minimum height=0.85cm,
        minimum width=4.8cm,
        align=center
    }
]


\node[box] (nbody)
{N-Body Simulations};

\node[box, below=0.45cm of nbody] (halo)
{Halo Catalogs};

\node[box, below=0.45cm of halo] (particle)
{Particle Maps};


\node[box, right=3cm of nbody] (tsz)
{AbacustSZ.py};

\node[box, below=0.45cm of tsz] (ygen)
{Compton-y Profile Generation};

\node[box, below=0.45cm of ygen, font=\bfseries] (ymap)
{Compton-y Map};


\draw
([xshift=-0.7cm]nbody.west)
--
([xshift=-0.7cm]particle.west);

\draw[-Stealth]
([xshift=-0.7cm]halo.west)
--
(halo.west);

\draw[-Stealth]
([xshift=-0.7cm]particle.west)
--
(particle.west);

\draw
([xshift=-0.7cm]nbody.west)
--
(nbody.west);


\draw[-Stealth] (halo.east) -- (ygen.west);

\draw[-Stealth] (particle.east) -- (ymap.west);

\draw[-Stealth] (tsz.south) -- (ygen.north);

\draw[-Stealth] (ygen.south) -- (ymap.north);


\node[
    below=0.4cm of particle,
    anchor=north,
    xshift=-0.2cm
] (countsimg)
{
\includegraphics[width=6.7cm]
{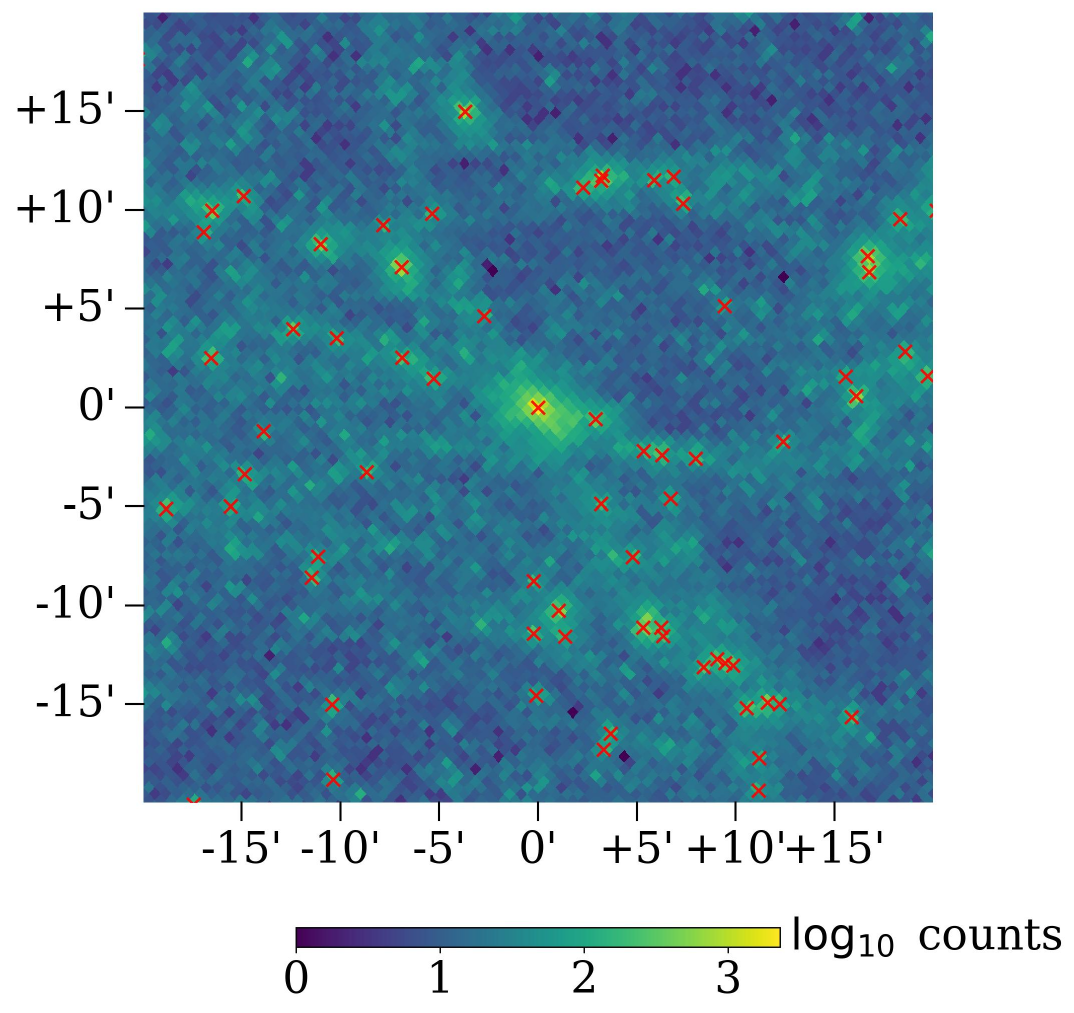}
};

\node[
    below=0.4cm of ymap,
    anchor=north,
    xshift=-0.4cm
] (yimg)
{
\includegraphics[width=5.8cm]
{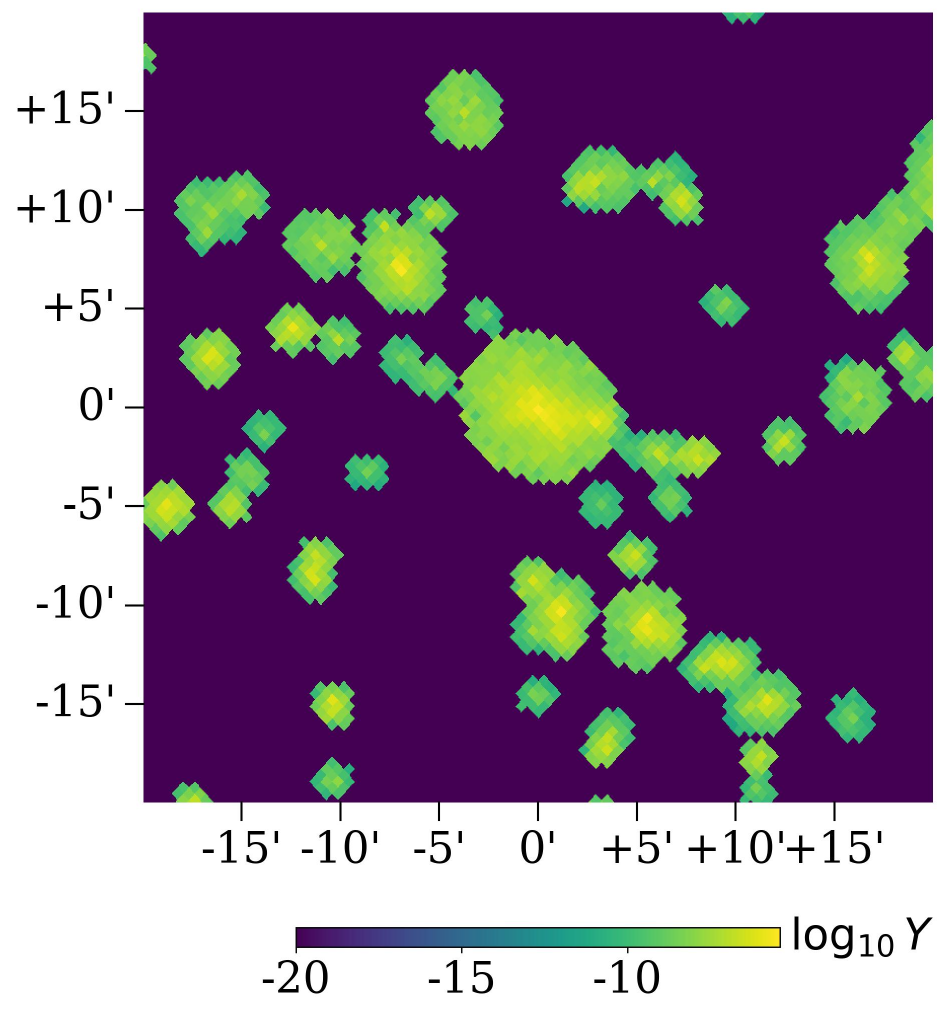}
};

\end{tikzpicture}%
}

\caption{
Flow diagram of the \texttt{AbacustSZ} pipeline. Halo catalogues and
\texttt{HEALPix} particle lightcone maps from the
\texttt{AbacusSummit} simulation suite are used as inputs. Using halo
masses, positions, and redshifts from the catalogues,
\texttt{AbacustSZ} paints projected Compton-$y$ profiles at the halo
positions and redistributes the signal within each halo using the
particle maps. The left panel shows a zoomed-in region of the
particle lightcone map at mean redshift $z=0.57$, centered at a halo of
mass $\log_{10}(M/M_\odot)=13.5$, together with marked halo centres
(red crosses). The right panel shows the resulting Compton-$y$ map
generated from the same sky region.
}

\label{fig:tszpaint_pipeline}

\end{figure*}

This section presents \texttt{AbacustSZ.py}, a simulation-based Python framework that combines analytic pressure-profile prescriptions with AbacusSummit simulation products to generate mock tSZ observables.
The pipeline is illustrated in Fig.~\ref{fig:tszpaint_pipeline}. Firstly, the projected Compton-$y$ profile is tabulated as a function of halo mass, redshift and angular separation for a given set of gNFW parameters (Sec.~\ref{sec:profile_generation}). Every light-cone halo is then painted onto a \texttt{HEALPix} map using this table (Sec.~\ref{sec:map_generation}). Within each halo, the painted signal is subsequently redistributed according to the local dark-matter distribution (Sec.~\ref{sec:signal_morph}). Summing the maps of all light-cone shells gives the full Compton-$y$ map.

\subsection{Simulation}

We use the \texttt{AbacusBacklight\_base\_c9999\_ph9999} simulation, which was run with the \textsc{Abacus} $N$-body code employed for the \texttt{AbacusSummit} suite \cite{Maksimova_2021,Garrison_2021}. The simulation evolves $2800^3\approx2.2\times10^{10}$ dark-matter particles in a periodic box of side $1\,h^{-1}$Gpc, with a particle mass of $m_p=4.0\times10^{9}\,h^{-1}M_\odot$, and provides full-sky light-cone halo catalogues alongside particle maps. This combination of volume and mass resolution is well suited for tSZ modelling, which is sensitive both to the abundance of rare massive halos and to the shape of halo pressure profiles. The simulation is part of the \texttt{Backlight} suite of full-sky simulations provided by Will Coulton, Lehman Garrison and Adrian Bayer. The adopted cosmological parameters are summarized in Table~\ref{tab:cosmology_parameters}.

\subsubsection{Halo catalogues}

The pressure-profile prescription employed by \texttt{AbacustSZ} requires information about the properties of individual halos. In \texttt{AbacusSummit}, halos are identified using the CompaSO halo finder \cite{10.1093/mnras/stab2980}, and we use the corresponding halo light-cone catalogues \cite{10.1093/mnras/stab3066}. The resulting catalogues provide halo positions, masses, and inertia-tensor eigenvalues and eigenvectors, which are used throughout the signal-generation procedure.

\textit{Halo masses.}---The pressure prescription of Sec.~\ref{sec:theory} is defined in terms of $M_{200c}$, whereas the CompaSO mass of an L1 halo, $N m_p \equiv M_{\rm vir}$, with $m_p$ denoting particle mass, is defined by a spherical-overdensity threshold relative to the mean matter density \cite{10.1093/mnras/stab2980}, which follows the virial overdensity of Bryan \& Norman \cite{BryanNorman1998}, $\Delta_{\rm L1}(z)=(200/18\pi^2)\,\Delta_{\rm vir,c}(z)/\Omega_m(z)$. We convert these masses to $M_{200m}$ and subsequently to $M_{200c}$ at the redshift of each light-cone shell, assuming an NFW profile \cite{Navarro1997UniversalProfile} with the concentration--mass relation of Diemer \& Joyce \cite{DiemerJoyce2019}, as implemented in \texttt{COLOSSUS} \cite{Diemer_2018}. The radius $R_{200c}$ then follows from $M_{200c}=\frac{4\pi}{3}\,200\,\rho_{\rm crit}(z)\,R_{200c}^3$.




\subsubsection{Particle lightcone maps}

\texttt{AbacusSummit} additionally provides particle light-cone maps in the equal-area \texttt{HEALPix} pixelisation \cite{Gorski_2005,Zonca2019}. Unlike the halo catalogues, which describe each halo through integrated properties, these maps retain the spatial distribution of the dark matter, which \texttt{AbacustSZ.py} uses to preserve the morphology of the tSZ signal (Sec.~\ref{sec:signal_morph}).

\vspace{0.5 cm}

\begin{table}[h]
\centering
\begin{tabular}{c c}
\hline
\textbf{Parameter} & \textbf{Value} \\
\hline
$\Omega_{m}$ & 0.3175 \\
$\Omega_{b}$ & 0.049 \\
$H_0$ & $67.11\ \mathrm{km\,s^{-1}\,Mpc^{-1}}$ \\
$\sigma_8$ & 0.834 \\
$n_s$ & 0.9624 \\
\hline
\end{tabular}
\caption{Cosmological parameters of the \texttt{AbacusBacklight\_base\_c9999\_ph9999} simulation used throughout this work. $\Omega_{m}$ is the total matter density, $\Omega_{b}$ -- baryon density fraction, $H_{0}$ -- Hubble constant, $\sigma_{8}$ -- amplitude of matter fluctuations on $8\,\mathrm{Mpc}/h$ scales, and $n_{s}$ -- spectral index of primordial fluctuations. The simulation assumes a flat $\Lambda$CDM cosmology.}
\label{tab:cosmology_parameters}
\end{table}


\subsection{Compton-$y$ Profile Generation\label{sec:profile_generation}}

In order to obtain the Compton-$y$ profiles, we require a numerical evaluation of the line-of-sight integral of Eq.~(\ref{eq:compton_y}), using the gNFW pressure prescription of Eq.~(\ref{eq:gnfw}). The resulting signal may therefore be written as $y(\theta\,|\,M_{200c}, z)$, where $\theta$ is the angular separation from the halo centre. As direct evaluation of this integral for each halo is computationally prohibitive, we employ an interpolation-based approach, inspired by the \texttt{XGPaint.jl} framework \cite{2025ascl.soft03033L}. The signal $y$ is precomputed on a regular grid in $(\ln\theta, z, \log_{10}M_{200c})$ spanning $-12<\ln(\theta/{\rm rad})<-1$, $0.2<z<1.0$ and $11<\log_{10}(M_{200c}/M_\odot)<15.7$ ($512\times128\times256$ points), and interpolated linearly in $\ln y$ for efficient capture of its large dynamic range. As the profile generation is intrinsically parallel, the tabulation has been implemented in JAX \cite{jax2018github}, with the line-of-sight integral evaluated by tanh--sinh quadrature using \texttt{quadax} \cite{Conlin_quadax}. This is crucial for the emulator of Sec.~\ref{sec:emulator}, where a new table must be constructed for every training point. Linear interpolation reproduces direct numerical integration to sub-percent accuracy, whilst higher-order schemes, such as cubic splines, were found to provide negligible improvements at significantly higher computational cost. 

\begin{figure}[t]
    \centering
    \includegraphics[width=\linewidth]{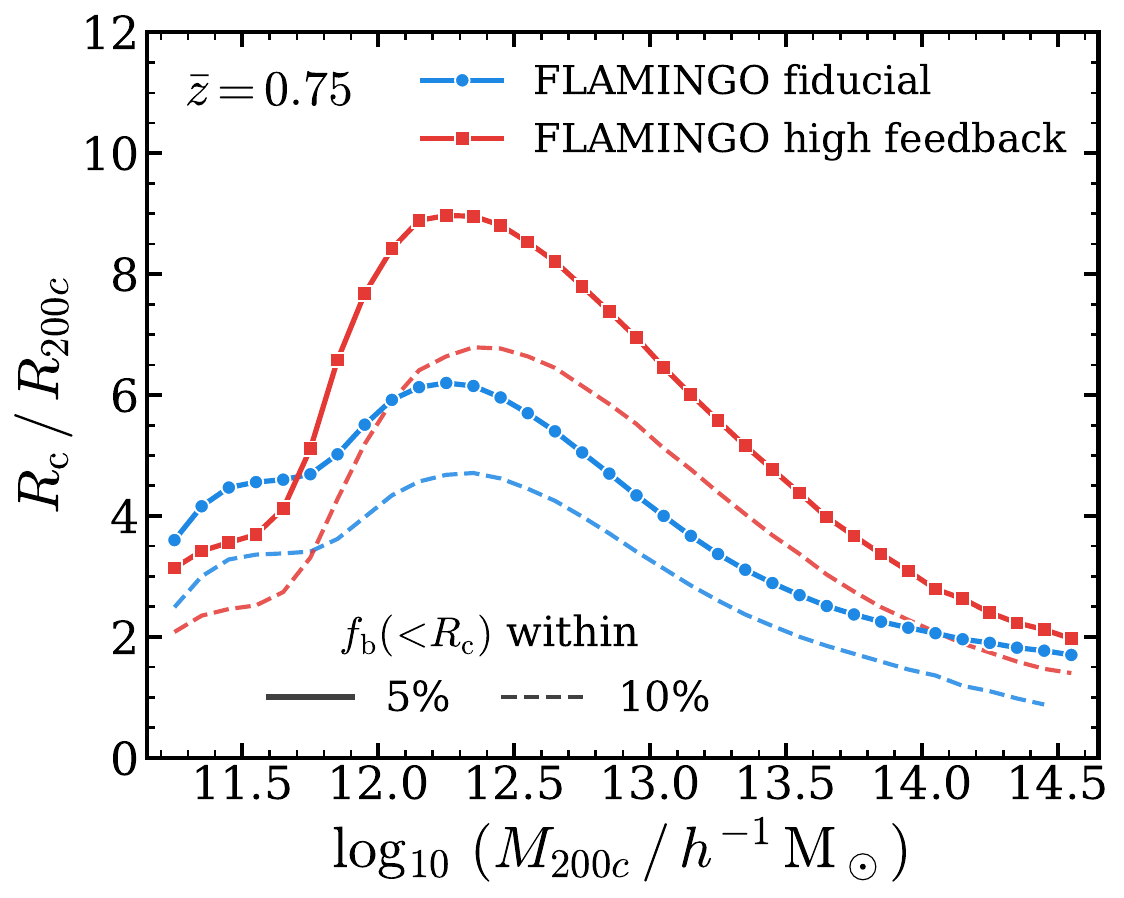}
    \caption{Closure radius $R_c$ in units of $R_{200c}$ as a function of halo mass, measured in the fiducial (blue circles) and high-feedback (red squares) FLAMINGO runs at $\bar z=0.75$. $R_c$ is the radius at which the enclosed baryon fraction first comes within 5\% (solid) or 10\% (dashed) of $\Omega_b/\Omega_m$.}
    \label{fig:closure_radius}
\end{figure}

\begin{figure*}[!t]
    \centering
    \includegraphics[width=\linewidth]{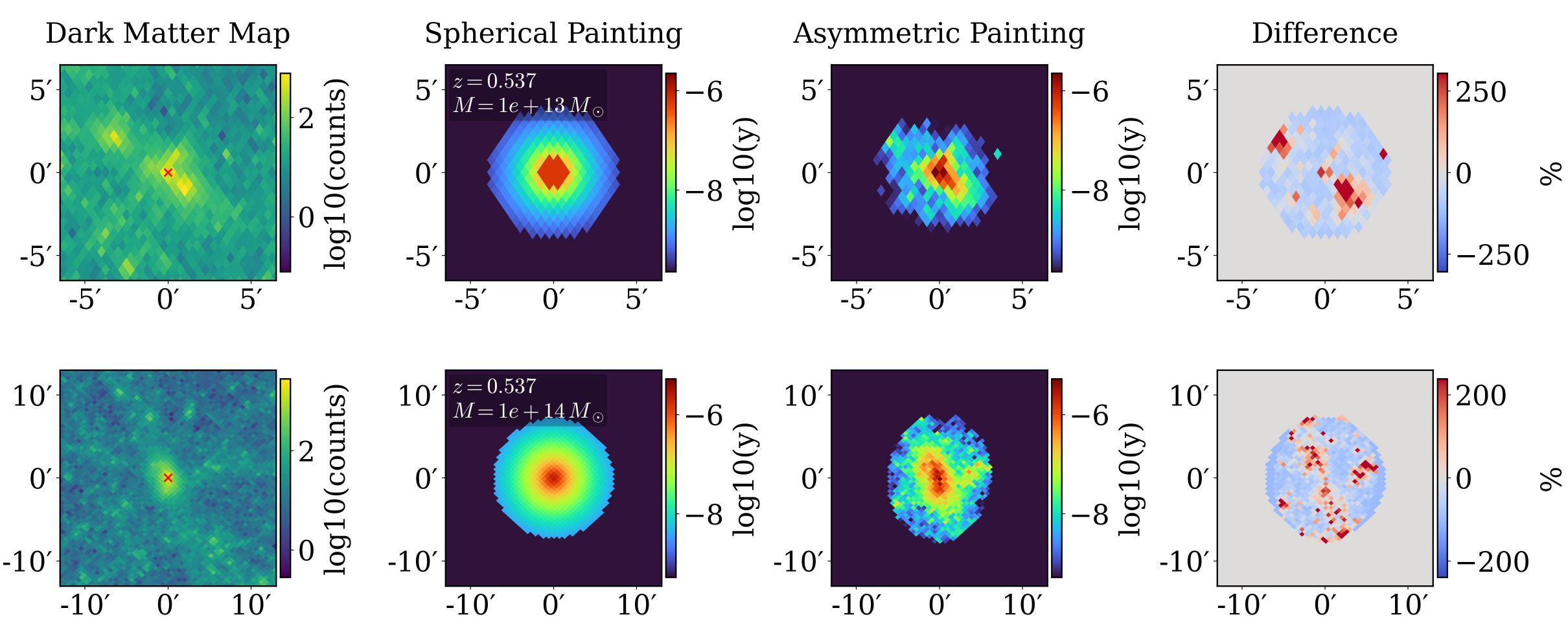}
    \caption{Comparison of the Compton-$y$ signals generated in \texttt{AbacustSZ.py} assuming a purely spherical pressure distribution and including the morphology-preserving signal redistribution scheme described in Section~\ref{sec:signal_morph}. The top row corresponds to a $M_{\rm{vir}} = 10^{13} M_\odot$ halo, while the bottom one to a $M_{\rm{vir}} = 10^{14} \ M_\odot$ one. The left column shows the underlying dark-matter particle maps for two example halos. The second column shows the corresponding projected Compton-$y$ signal obtained assuming spherical symmetry, and the third column shows the generated signal with retained morphological features. The last column visualizes the spatial difference between the signal contribution of the two prescriptions.}
    \label{fig:visualization}
\end{figure*}

\textit{Truncation at the closure radius.}---The extent to which the pressure profile is integrated is a modelling choice. For concentrated, quickly-converging pressure profiles, such as the fiducial Battaglia et al. \cite{Battaglia_2012}, the cut-off does not influence the overall signal. However for shallow pressure profiles, outskirts keep contributing, and the truncation radius serves as a physical cut-off of where the gas extends to. Map-making codes typically truncate the gNFW profile at a fixed multiple of the halo radius, e.g.\ $4R_{200c}$ in \texttt{XGPaint.jl} \cite{2025ascl.soft03033L}, $5R_{200c}$ in Ref.~\cite{Kovac2026}, or around $\approx2R_{200c}$ based on the calibration from Ref.~\cite{Battaglia_2012}.
We instead adopt a physically motivated, mass-dependent truncation at the baryon closure radius $R_c$, defined as the halocentric radius within which the cumulative baryon fraction reaches the cosmic value $\Omega_b/\Omega_m$ \cite{Ayromlou_2023}. We adopt the radius at which it first comes within 5\% of this value. Following the method of Ayromlou et al.~\cite{Ayromlou_2023}, we measure the closure radius $R_c(M_{200c})$ in the high-feedback ($f_{\rm gas}-8\sigma$) and fiducial runs of the FLAMINGO simulations \cite{Schaye2023FLAMINGO,Kugel2023FLAMINGO}, as illustrated in Figure~\ref{fig:closure_radius}. Physically, $R_c$ marks the radius out to which a halo has redistributed its gas: feedback expels baryons from the halo centre, but within $R_c$ the halo again contains its full cosmic share of baryons. Truncating at $R_c$ therefore retains the gas displaced by feedback, without extending each profile indefinitely. We choose the high-feedback run for our later forward model analysis. This serves as a limiting case and ensures no gas is omitted from ``painting'': if the gas pressure profile were less extended than the high-feedback scenario predicts, then the fitted gNFW would simply contribute negligibly near $R_{c}$. This truncation is applied in all dimensions, both in this interpolated line-of-sight integral, as well as for pixel-halo assignments described in Sec.~\ref{sec:map_generation}.





\subsection{Compton-$y$ Map Generation\label{sec:map_generation}}

Firstly, halo positions and masses are extracted from the light-cone halo catalogues. For each halo, the pixels lying within a projected radius $R_c(M_{200c})$ of its centre, and within the elliptical footprint of Sec.~\ref{sec:signal_morph}, are identified, and each pixel is assigned the interpolated $y$ at its angular separation from the halo centre. 

\textit{Overlapping halos.}---With truncation at $R_c$, the painted volumes of neighbouring halos frequently overlap, with about 85\% of the painted halos intersecting at least one neighbour. To avoid double-counting of the signal, we use competitive assignment of every point in space to a single halo, namely the one closest to it in units of its own radius.
For each pixel--halo pair, the line of sight crosses the painted sphere along a chord of length $2\sqrt{R_{c}^{2} - b^{2}}$, where $b$ is the impact parameter. We compute the fraction of the pressure integral along this chord which lies in the region the halo owns through a precomputed table of cumulative gNFW chord integrals, which are functions of $b/R_c$, halo mass and position along the chord, and multiply the painted $y$ by this fraction. 

The fiducial settings paint the tSZ signal for all halos with $N m_p \equiv M_{\rm vir} >10^{11.5}\,h^{-1}M_\odot$ (about 80 particles). This reflects both the strong weighting of the tSZ signal towards massive halos and the rapidly increasing computational cost of including lower-mass halos.

\subsection{Signal morphology \label{sec:signal_morph}}

In \texttt{AbacustSZ}, we preserve the asymmetric projected tSZ morphology, whilst retaining the underlying radial dependence of the gNFW pressure profile. Using the provided dark-matter particle lightcone maps, and treating particle counts as a proxy for the underlying matter density field, we redistribute the signal according to the local particle distribution. Within each radial annulus around the halo centre, $\theta_i < \theta < \theta_{i+1}$ (of width $2\times10^{-5}$\,rad, or at least one pixel), the projected Compton-$y$ signal is redistributed across the enclosed pixels using weights $w_p = N_{\mathrm{counts}}^{5/3}$, where $N_{\mathrm{counts}}$ denotes the number of particles in each pixel, motivated by the approximate scaling relation $Y \propto M^{5/3}$.
Pixels containing more dark matter receive a larger share of the signal (Fig.~\ref{fig:visualization}). The Compton-$y$ assigned to pixel $p$ in annulus $i$ is then
\begin{equation}
\label{eq:redistribution}
y_p = y(\theta_p)\,\frac{N_i\, w_p}{\sum_{p'\in i} w_{p'}},
\end{equation}
where $y(\theta_p)$ is the interpolated profile at the pixel's angular separation and $N_i$ is the number of pixels in the annulus, so that the mean weight within each annulus is unity.

\begin{figure*}[t]
    \centering
    \includegraphics[width=\linewidth]{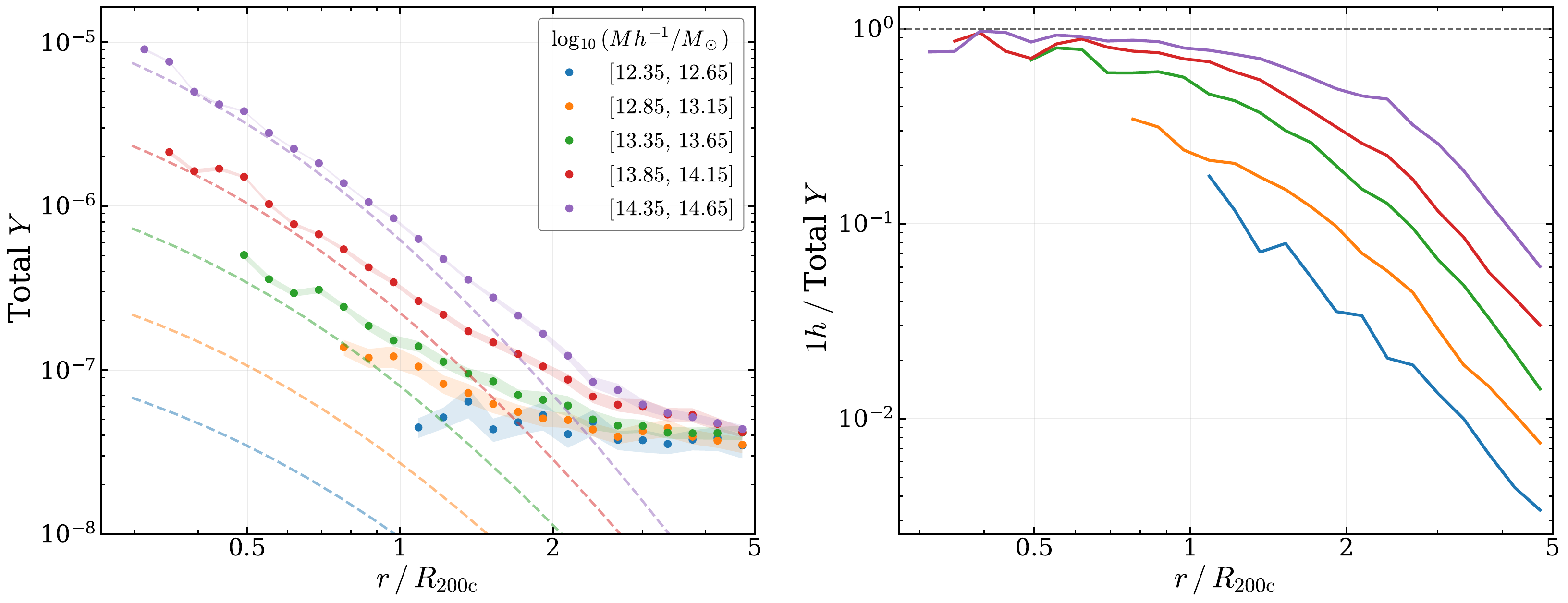}
    \caption{Compton-$y$ profiles generated in \texttt{AbacustSZ.py} and the ratio between one-halo and total signal, painted for the fiducial Battaglia et al.(2012) profile \cite{Battaglia_2012}. \textit{Left:} Simulated Compton-$y$ signal as a function of projected distance from the halo centre, normalized by $R_{200c}$, for different halo-mass bins. The mean stacked signal for 1000 halos in each mass bin is evaluated through radial binning (points), together with the corresponding analytical one-halo profile prediction (dashed lines). \textit{Right:} Ratio between the analytical one-halo contribution and the total simulated signal.  Beyond $r \sim R_{200c}$, the two-halo term ($\mathrm{total} - \text{1-halo}$) begins to dominate across all mass bins. }
    \label{fig:profiles_two_halo}
\end{figure*}


 Dark matter halos are generally triaxial rather than spherical \cite{Jing2002}, and AbacusSummit halo shapes are characterised by the given inertia-tensor eigenvectors and eigenvalues. We use this information to construct an elliptical footprint for each halo on the sky (see Appendix~\ref{app:halo_shapes} for derivation), and query only the pixels inside it. The semi-axes are normalized such that $abc=R_{200c}^3$, preserving the volume of the spherical halo, and then scaled by $R_c/R_{200c}$ so that the painted region extends to the closure radius. Only the footprint is elliptical: the value assigned to each pixel is still evaluated at its angular separation from the halo centre.

A fully triaxial pressure model was not implemented, as this would require evaluating the line-of-sight integral as an explicit function of halo orientation and axis ratios, substantially increasing the dimensionality and computational cost of the interpolation procedure. The adopted implementation instead provides a computationally efficient approximation that preserves the dominant projected morphological features of the tSZ signal.


\subsection{Two-halo term in \texttt{AbacustSZ}}

The two-halo term arises within \texttt{AbacustSZ.py} from the clustering of halos in the AbacusSummit light cone. Figure~\ref{fig:profiles_two_halo} shows the generated Compton-$y$ signal as a function of projected distance from the halo centre normalized by $R_{200c}$, together with the relative contribution of the 2-halo term to the total signal. The 2-halo contribution becomes increasingly significant for lower-mass halos and dominates across much of the radial range, consistent with previous findings in \cite{Hill_2018}. In the halo model (Appendix~\ref{app:halo_model}), the one-halo term follows approximately the self-similar scaling $Y \propto M^{5/3}$, whereas the two-halo term scales with the halo bias, which depends much more weakly on mass \cite{Hill_2018,Tinker2010}. 

The decreasing number of radial data points with decreasing halo mass in Fig.~\ref{fig:profiles_two_halo} arises from the finite pixel resolution of the \texttt{HEALPix} maps, with $R_{200c}$ of the lowest-mass halos approaching the characteristic pixel scale of $\approx0.43$\,arcmin at $N_{\rm side}=8192$. This is however comfortably below the $\approx$ 1.5 arcmin FWHM of the beams used in typical analyses \cite{Naess_2020}.

For completeness, the recovery of the analytical gNFW one-halo profile within \texttt{AbacustSZ.py} was additionally validated prior to projection onto the \texttt{HEALPix} maps, following the same procedure as in Fig.~\ref{fig:profiles_two_halo}. Agreement at the level of better than $0.5\%$ was obtained across all radial bins.

\section{Modelling ACT tSZ $\times$ DESI LRG measurements\label{sec:modelling}}

We now aim to constrain the gas pressure profile using the Atacama Cosmology Telescope (ACT) Data Release 5 (DR5) $\times$ DESI LRG tSZ measurement presented in  Ref.~\cite{liu2025measurementsthermalsunyaevzeldovicheffect}.  We first describe the measurement and the forward model, and then present the resulting constraints on the pressure profile.

\subsection{Data}

The modelled stacked tSZ measurements in this work combine the DESI photometric LRG catalogue \cite{Dey2019,Zhou2023LRGTargets} divided into four photometric-redshift bins \cite{Zhou_2023}, with microwave temperature maps from the ACT DR5 \cite{Naess_2020}.

A major challenge in tSZ analyses is contamination from astrophysical foregrounds, most notably emission from the Cosmic Infrared Background (CIB). In this work, we use the stacks on the single-frequency ACT DR5 maps in the f090 (77--112\,GHz) and f150 (124--172\,GHz) bands, presented in Appendix~C of Ref.~\cite{liu2025measurementsthermalsunyaevzeldovicheffect}, as they proved to be less prone to residual dust contamination than the fiducial stacks on component-separated Compton-$y$ maps \cite{PhysRevD.109.063530}. Among the four DESI redshift bins, we select the second bin for analysis, corresponding to a mean redshift of $z_{\rm mean}=0.628$. Potential dust contamination in this bin was estimated using the ACT f220 map, where the tSZ signal is small, and was found to be negligible on large scales for this bin \cite{liu2025measurementsthermalsunyaevzeldovicheffect}. Furthermore, the raw (not dust-corrected) f090 and f150 ring-ring profiles (see definition of the ``ring-ring'' filter in Eq.~\ref{eq:ringring}) of this bin are already in good agreement (Fig.~14 of Ref.~\cite{liu2025measurementsthermalsunyaevzeldovicheffect}). As the tSZ and dust signals scale differently between the two bands, this provides additional evidence that residual foreground contamination is subdominant. We fit only the f090 profile, for two reasons. For this bin, its per-aperture uncertainties are 6--12\%, compared with 10--23\% at f150, and thermal dust emission, the main contaminant, is substantially weaker at 90\,GHz than at 150\,GHz.

\begin{figure}[h]
    \centering
    \includegraphics[width=0.9\linewidth]{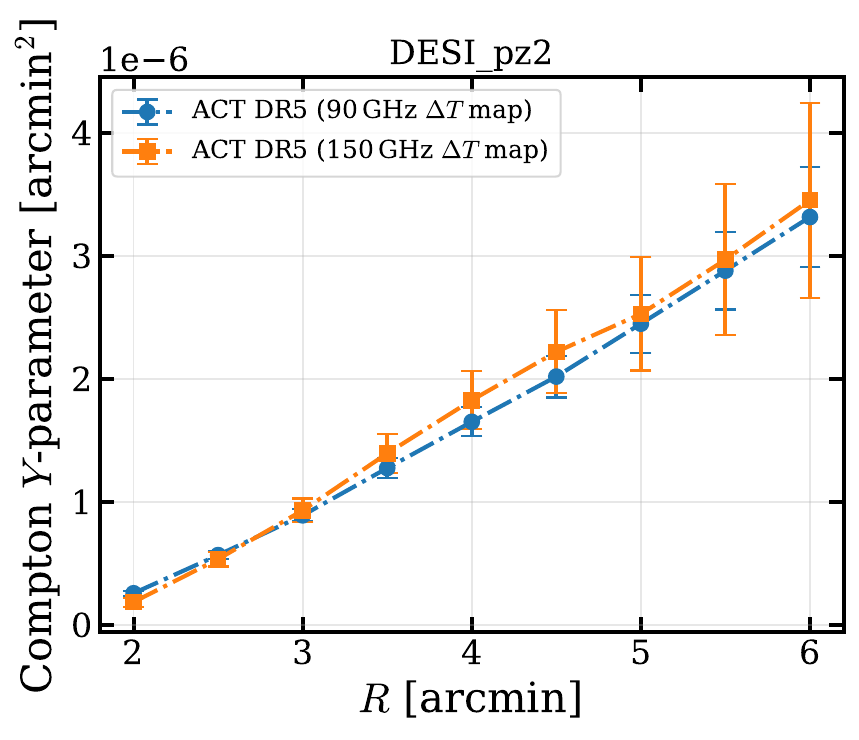}
    \caption{Stacked ACT DR5 ring-ring profiles of redshift bin 2, from Fig.~15 of Ref.~\cite{liu2025measurementsthermalsunyaevzeldovicheffect}. The agreement between the two single-frequency measurements suggests that any unaccounted-for frequency-dependent component is unlikely to significantly contaminate the measurement. Since the two single-frequency maps have a different effective beam, this also indicates that there may be some small vestigial dust contamination in the first radial bin. 
    We fit the f090 profile.}
    \label{fig:data}
\end{figure}

\begin{figure}[h]
    \centering
    \includegraphics[width=1\linewidth]{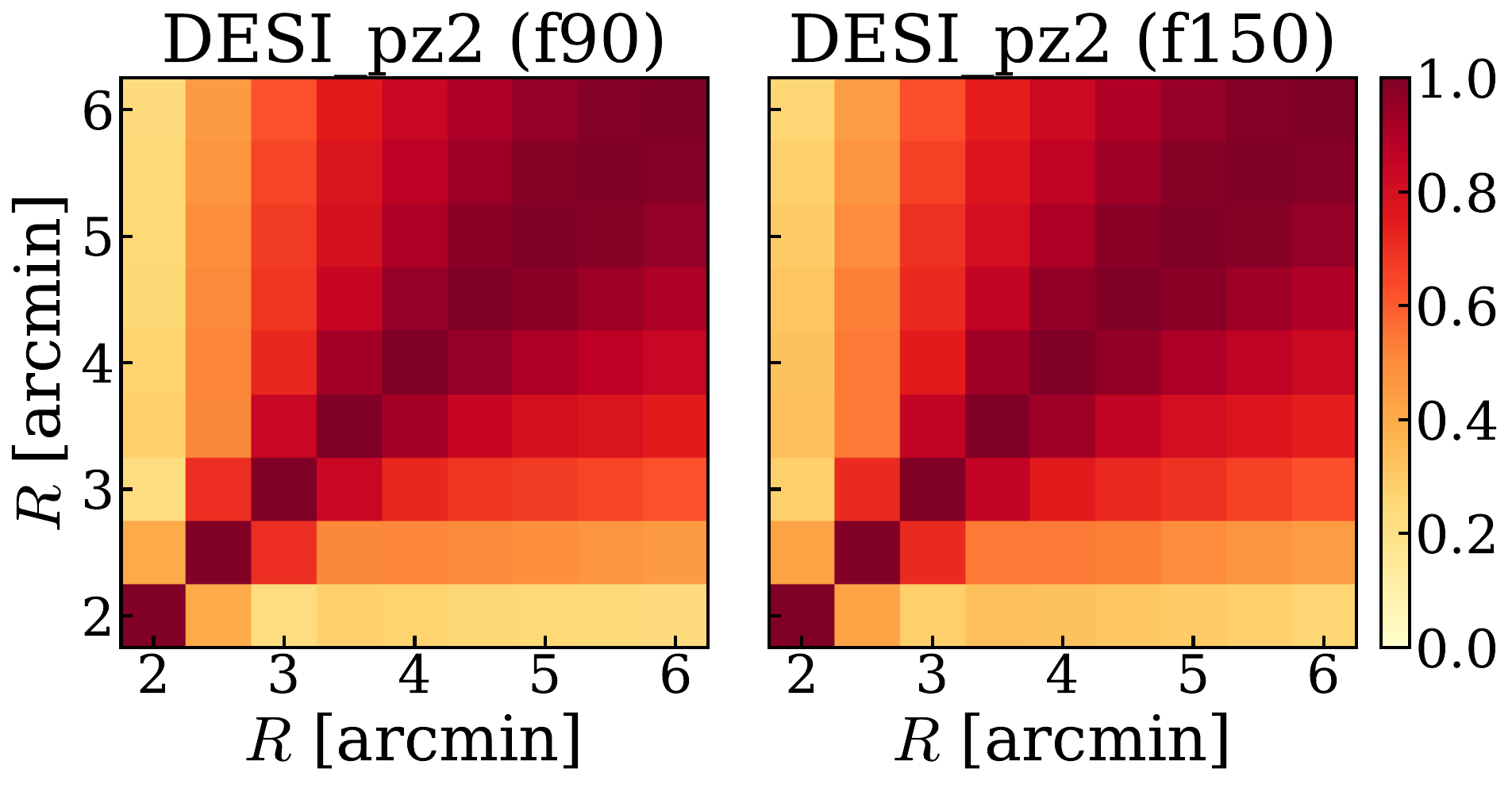}
    \caption{Correlation matrices of the stacked profiles of Fig.~\ref{fig:data} \cite{liu2025measurementsthermalsunyaevzeldovicheffect}. At large apertures (beyond 3--4$'$), there is large correlation between the radial bins, indicating that the majority of the constraining power resides in the smallest apertures ($\lesssim 3'$).}
    \label{fig:covariance}
\end{figure}

The stacked profiles of this bin and their correlation matrices are shown in Figs.~\ref{fig:data} and \ref{fig:covariance}.





\subsubsection{Stacking Procedure}

The stacked tSZ ACT$\times$ DESI measurements are constructed using a Compensated Aperture Photometry (CAP). CAP filtering is widely used in SZ analyses because it suppresses large-scale fluctuations, such as primary CMB anisotropies and slowly-varying foreground emission \cite{Schaan_2021}. For a galaxy at angular position $\boldsymbol{\theta}_i$, the filtered Compton-$y$ signal at aperture radius $\theta_d$ is
\begin{equation}
T_i(\theta_d)
= \int d^{2}\boldsymbol{\theta}\; y(\boldsymbol{\theta}_i+\boldsymbol{\theta})\, W_{\theta_{d}}(|\boldsymbol{\theta}|),
\end{equation}
where $W_{\theta_d}$ is the CAP window function, so that $T_i$ has units of ${\rm arcmin}^2$. Physically, the filter measures the average signal within a central aperture and subtracts the average signal from a surrounding annulus of equal area. By construction, the filter integrates to zero.

The stacked profile is then obtained by averaging the filtered measurements over all galaxies in the sample,

\begin{equation}
T(\theta_d)
=
\frac{1}{N}
\sum_i T_i(\theta_d),
\end{equation}

where $N$ is the total number of galaxies. The ACT$\times$DESI analysis further introduces the ring-ring filter, corresponding to the semi-cumulative window function

\begin{equation}
\label{eq:ringring}
\tilde{W}_{\theta_0,\theta_d}(\theta)=
\begin{cases}
1 & \theta_0 < \theta < \theta_d, \\
-1 & \theta_d < \theta < \sqrt{2\theta_d^2-\theta_0^2}, \\
0 & \text{otherwise}.
\end{cases}
\end{equation}

Here, $\theta_0 = 1'$ defines an inner masking radius, which excludes the innermost region of the galaxy (hence, ring-ring). This was adopted to suppress contamination from dusty star-forming galaxies and the associated CIB emission, which is expected to be most significant near the galaxy centre \cite{liu2025measurementsthermalsunyaevzeldovicheffect}.

\subsection{Forward model\label{sec:forward_model}}

We vary four gNFW parameters, $\boldsymbol{\theta}=(\alpha,\beta_0,\gamma,\log_{10}P_0)$ (Eq.~\ref{eq:gnfw_scaling}), with uniform priors over the ranges listed in Table~\ref{tab:gnfw_ranges}.
These ranges were chosen to comfortably contain the region favoured by a preliminary semi-analytical fit to the same measurements (Appendix~\ref{app:halo_model}). We fix $x_c$ to its fiducial Battaglia et al. value \cite{Battaglia_2012}, because of its strong degeneracies with the remaining gNFW parameters, and additionally keep the mass and redshift exponents of Eq.~(\ref{eq:gnfw_scaling}) at their fiducial values. Each evaluation of the forward model requires painting new tSZ maps and stacking them around millions of mock galaxies, producing over 500\,GB of intermediate data products, which makes direct likelihood evaluation computationally prohibitive. We therefore evaluate the forward model on a fixed training design, and interpolate between the design points with a Gaussian-process (GP) emulator (Sec.~\ref{sec:emulator}).

\subsubsection{Forward Model Pipeline}

For each gNFW variation, we construct the Compton-$y$ interpolators using \texttt{AbacustSZ.py}, paint the 20 light-cone shells spanning $0.49<z<0.79$, corresponding to the redshift range of the LRG mocks, and sum them along the line of sight into a single map. In our case, painting circular instead of elliptical footprints, does not change the stacked profile appreciably, as the stack averages over many halo orientations, hence spherical implementation was adopted.

The summed map is subsequently convolved with a Gaussian beam of FWHM 2.1\,arcmin, matching the resolution of the ACT f090 map \cite{Naess_2020}.
The ACT$\times$DESI analysis additionally masks SZ-detected clusters \cite{liu2025measurementsthermalsunyaevzeldovicheffect}. We approximate this masking by not ``painting'' halos above $M_{500c}=3.8\times10^{14}\,M_\odot$, corresponding to the 90\% completeness limit of the ACT DR5 cluster catalogue at $z=0.5$ \cite{Hilton2021}, converted to $M_{200c}$ using the same NFW and concentration assumptions as above.

Mock LRGs are populated using the \texttt{AbacusHOD} package \cite{Yuan_2021}, which employs the Halo Occupation Distribution (HOD) model of Zheng et al.~\cite{Zheng_2007}. We adopt the best-fitting HOD parameters of the DESI DR2 spectroscopic LRGs used in Ref.~\cite{Qu2026}, interpolated linearly in redshift between their fits at $z=0.50$ and $z=0.725$ to the mean redshift of our sample, $z=0.628$: $\log M_{\rm cut}=12.66$, which sets the characteristic halo mass for central-galaxy occupation; $\log M_1=13.79$, the characteristic mass scale for satellite occupation; $\alpha=1.07$, the power-law slope governing the growth of satellite occupation with halo mass, and $\kappa = 0.92$, which sets the minimum halo mass for satellites. Masses follow the virial definition in the COMPASO halo finder, and are given in units of $h^{-1}M_\odot$. As the spectroscopic LRGs are slightly more massive than the photometric sample considered here, we randomly subsample the most massive host halos, such that the mean host mass matches $\log_{10}(\bar{M}_{200c}/h^{-1}M_\odot)=13.19$ measured from CMB lensing for the same photometric sample within its' $0.1 \rm dex$ uncertainty \cite{Hadzhiyska2025lensing}. Matching the mean host mass is the leading requirement, as it sets the typical halo size and tSZ amplitude around the stacked galaxies, which scales with $y \propto M^{5/3}$. The sensitivity to the remaining HOD freedom is discussed in Sec.~\ref{sec:hod} and will be the subject of further work.

Galaxies hosted by halos above the same mass threshold are removed, and the catalogue is randomly downsampled to $3\times10^{6}$ galaxies for computational ease, ensuring that the statistical error resulting from this operation is much smaller than that of the observational data ($ <5 \%$). The signal around each galaxy is then measured by applying the ring-ring aperture filter at nine apertures, $2\le\theta_d\le6$\,arcmin and averaged over the sample. The resulting stacked radial profiles form the training set for the GP emulator.






\begin{figure*}[t!]
    \centering

    \includegraphics[width=\linewidth]{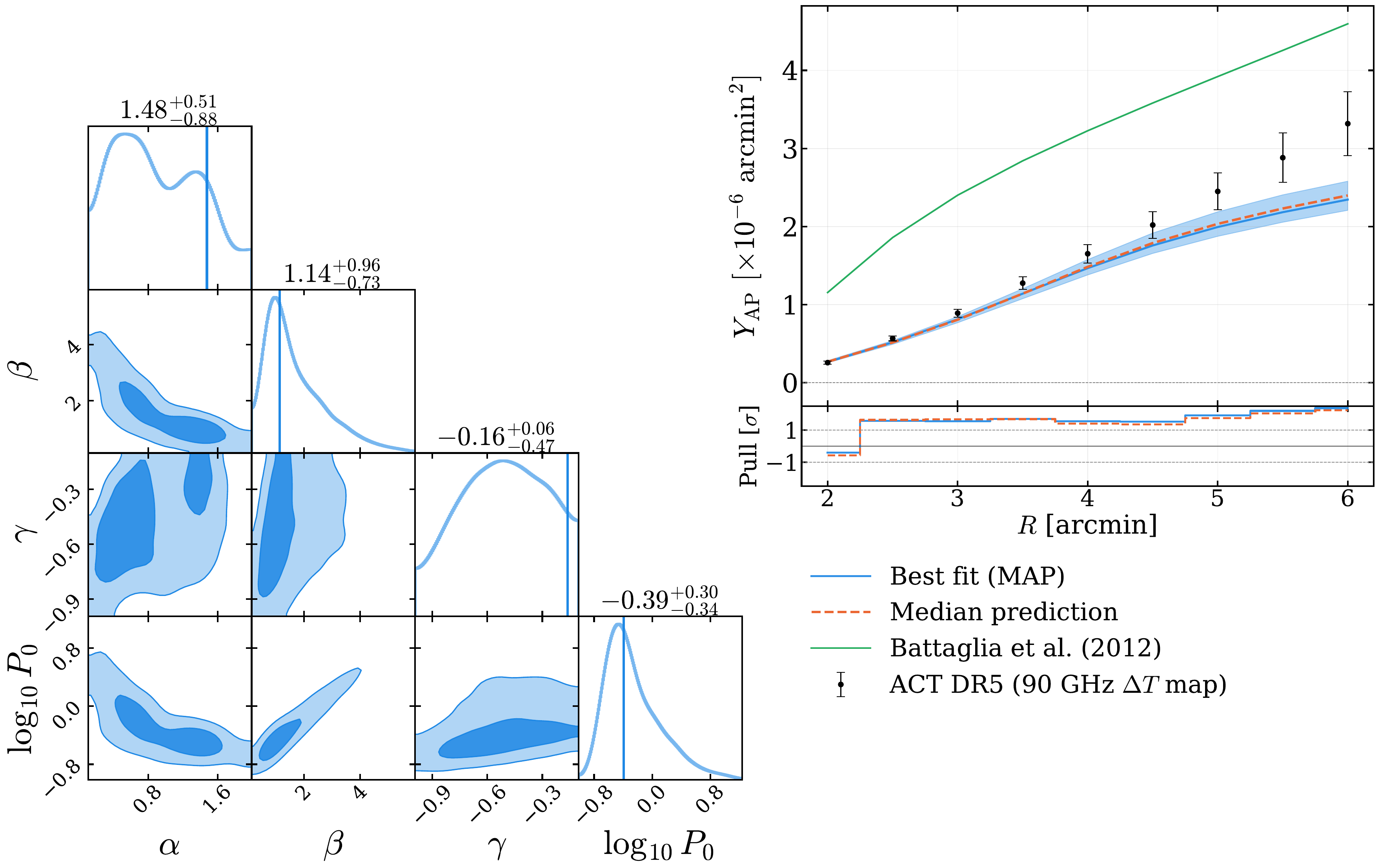}

    \caption{
    Posterior constraints on the gNFW pressure-profile parameters and the
    corresponding best-fitting stacked tSZ profile. The corner plot shows
    the marginalized posterior distributions of the transition sharpness
    $\alpha$, fall-off exponent $\beta_0$, inner slope $\gamma$, and normalization
    $\log_{10}P_0$ inferred from the stacked tSZ measurements. The profile
    panel shows the maximum-a-posteriori (MAP) prediction (blue line) and the posterior median (red dashed line),
    together with the corresponding $68\%$ credible interval (blue shaded
    region) and the normalized residuals,
    $(\mathrm{data}-\mathrm{model})/\sigma_{\mathrm{data}}$. The prediction
    obtained using the fiducial Battaglia et al.~(2012) gNFW parameter set
    is shown in green for comparison \cite{Battaglia_2012}.
    }

    \label{fig:gnfw_constraints}
\end{figure*}

\begin{table*}[t!]
    \centering

    \renewcommand{\arraystretch}{1.25}
    \setlength{\tabcolsep}{10pt}

    \begin{tabular}{cccc}
        \hline
        \textbf{Parameter} &
        \textbf{Description} &
        \textbf{Prior range} &
        \textbf{$1\sigma$ Constraints} \\
        \hline

        $\alpha$ &
        Transition sharpness &
        $\mathcal{U}[0.1,\,2]$ &
        $1.48^{+0.51}_{-0.88}$ \\

        $\beta_0$ &
        Fall-off exponent &
        $\mathcal{U}[0.1,\,6]$ &
        $1.14^{+0.96}_{-0.73}$ \\

        $\gamma$ &
        Inner slope &
        $\mathcal{U}[-1,\,-0.1]$ &
        $-0.16^{+0.06}_{-0.47}$ \\

        $\log_{10} P_0$ &
        Pressure-profile normalization &
        $\mathcal{U}[-2,\,2]$ &
        $-0.39^{+0.30}_{-0.34}$ \\

        \hline
    \end{tabular}

    \caption{Free gNFW parameters of the fit, their uniform priors and marginalized constraints (median and 16th--84th percentiles). $\beta_0$ and $P_0$ are the values of $\beta$ and $P_0$ at $M_{200c}=10^{14}\,M_\odot$ and $z=0$ [Eq.~(\ref{eq:gnfw_scaling})].}
    \label{tab:gnfw_ranges}

\end{table*}

\subsubsection{Gaussian-process emulator\label{sec:emulator}}

The training design consists of 130 points, drawn by Latin-hypercube sampling \cite{McKay1979} of the four-dimensional prior volume of Table~\ref{tab:gnfw_ranges}, together with the fiducial parameters of Battaglia et al. \cite{Battaglia_2012} (Table~\ref{tab:battaglia_parameters}). 

The emulator exploits the fact that the stacked profile is exactly proportional to $P_0$, as the closure radii, ownership fractions, cluster cut, beam convolution and aperture filter depend only on the profile shape and on the halo masses, and not on the pressure normalization. We therefore emulate $\log_{10}[T(\theta_d)/P_0]$ as a function of $(\alpha,\beta_0,\gamma)$ only, and restore $P_0$ analytically. This removes one dimension from the interpolation, and the projection of the design onto $(\alpha,\beta_0,\gamma)$ retains the one-dimensional stratification of the Latin hypercube in each parameter.

For computational simplicity, we model each of the nine apertures as an independent GP \cite{Rasmussen2005}, implemented in \texttt{scikit-learn} \cite{scikit-learn}. We use a Mat\'ern-5/2 kernel with a separate length scale for each parameter, as it performed best in hold-out tests among standard smooth kernels. The inputs are standardized to zero mean and unit variance, and the targets separately for each aperture. The kernel amplitude and length scales are fitted to the training profiles. The GP predictive standard deviation, propagated to $T(\theta_d)$, then provides the emulator uncertainty $\sigma_{\rm GP}(\boldsymbol{\theta})$ entering the likelihood. 

We validate the emulator by leave-one-out cross-validation \cite{Rasmussen2005}. Each training point is removed in turn, the emulator is retrained on the remaining points, and the fractional error $(m-T)/T$ of the predicted profile is recorded at each aperture. The median absolute error is $0.4\%$, and $95\%$ of the predictions lie within $5\%$ of the true profile. This is comparable to the emulator's own predicted $\sigma_{GP}$, on the order of $0.5\%$, which is subsequently propagated into the covariance matrix for inference \ref{sec:inference}. The emulator error is  well below the 6--12\% uncertainties of the measurement.





\subsubsection{Bayesian inference\label{sec:inference}}

The emulator prediction $\mathbf{m}(\boldsymbol{\theta})$ is compared with the measured f090 profile $\mathbf{d}$ through the Gaussian likelihood
\begin{equation}
\label{eq:likelihood}
-2\ln\mathcal{L}(\boldsymbol{\theta}) = \mathbf{r}^{\rm T}\mathsf{C}^{-1}\mathbf{r} + \ln\det\left(2\pi\mathsf{C}\right),
\end{equation}
with residual $\mathbf{r}=\mathbf{d}-\mathbf{m}(\boldsymbol{\theta})$ and total covariance $\mathsf{C}(\boldsymbol{\theta})=\mathsf{C}_{\rm data}+{\rm diag}[\sigma_{\rm GP}^2(\boldsymbol{\theta})]$, which combines the observational covariance matrix with the predicted emulator uncertainty. The determinant term is retained, as the emulator uncertainty depends on $\boldsymbol{\theta}$. For $\mathsf{C}_{\rm data}$, we take the correlation matrix of the measurement (Fig.~\ref{fig:covariance}) and rescale it by the published per-aperture uncertainties \cite{liu2025measurementsthermalsunyaevzeldovicheffect}. Since the apertures are emulated independently, correlations between their emulator errors are neglected.

The posterior distribution over the prior range is then sampled using the \texttt{dynesty} nested-sampling algorithm \cite{2004AIPC..735..395S,Skilling2006,2020MNRAS.493.3132S,Koposov_dynesty_zenodo}. We adopt 4000 live points, multi-ellipsoid bounds \cite{Feroz2009} and random-slice sampling \cite{Handley2015a,Handley2015b} with 20 slices, together with a termination criterion of $\Delta\ln Z<0.01$ on the estimated remaining evidence. We quote the marginalized medians with 16th--84th percentile intervals, alongside the maximum a posteriori (MAP) point, which for uniform priors corresponds to the maximum-likelihood point within the prior volume. The goodness of fit is quantified by $\chi^2=\mathbf{r}^{\rm T}\mathsf{C}^{-1}\mathbf{r}$, evaluated at the MAP point. As weakly constrained and degenerate parameters do not each remove a full degree of freedom \cite{andrae2010dosdontsreducedchisquared}, we evaluate its probability to exceed (PTE) for $N_d-N_{\rm eff}$ degrees of freedom, with $N_d=9$ apertures and $N_{\rm eff}=\langle\chi^2\rangle_{\rm post}-\chi^2_{\rm MAP}$ the Bayesian complexity \cite{Kunz2006}.

\subsection{Results\label{sec:results}}

\subsubsection{Pressure Profile Constraints}

The posterior distributions and best-fitting stacked profile are shown in Fig.~\ref{fig:gnfw_constraints}, with marginalized parameter constraints summarized in Table~\ref{tab:gnfw_ranges}.
We obtain $\chi^2 = 11.77$ at the MAP point and $N_{\rm eff}=1.9$, which reflects the weak constraints on $\alpha$ and $\gamma$ and the degeneracy between $\beta_0$ and $\log_{10}P_0$. For the resulting 7.1 degrees of freedom, this gives ${\rm PTE}=0.11$, indicating an acceptable fit to the data. This PTE is furthermore likely conservative, as the covariance of Ref.~\cite{liu2025measurementsthermalsunyaevzeldovicheffect} is estimated by bootstrap resampling of individual galaxies, which assumes that the noise is independent from galaxy to galaxy. Since the filters of neighbouring galaxies overlap on the sky, this assumption does not strictly hold \cite{liu2025measurementsthermalsunyaevzeldovicheffect}, and the uncertainties may be underestimated compared with a covariance estimated by resampling spatial blocks. The best-fit profile reproduces the innermost aperture, which carries the smallest uncertainty, whilst it slightly underpredicts the data at the largest apertures, where the two-halo term becomes increasingly important.

The inferred pressure profile favours low values $\beta_0$, and a gradual transition and a shallow outer logarithmic slope $\gamma-\alpha\beta$. The pressure therefore falls off more slowly with radius than in the fiducial model, placing more of the signal at large radii. For comparison, Fig.~\ref{fig:gnfw_constraints} also shows the stacked profile predicted using the fiducial Battaglia et al.~(2012) gNFW calibration \cite{Battaglia_2012}, which significantly overpredicts the overall signal, and does not reproduce a shape consistent with the data. It is important to note, however, that this fiducial calibration is an extrapolation at the LRG masses analysed in this analysis, as the Battaglia et al.~\cite{Battaglia_2012} profile was calibrated on massive simulated clusters. As a cross-check, Appendix~\ref{app:halo_model} fits the same measurement with a semi-analytic halo model, in which the amplitude of the two-halo term is a free parameter.

\begin{figure}[h!] 
\centering 
\includegraphics[width=1\linewidth]{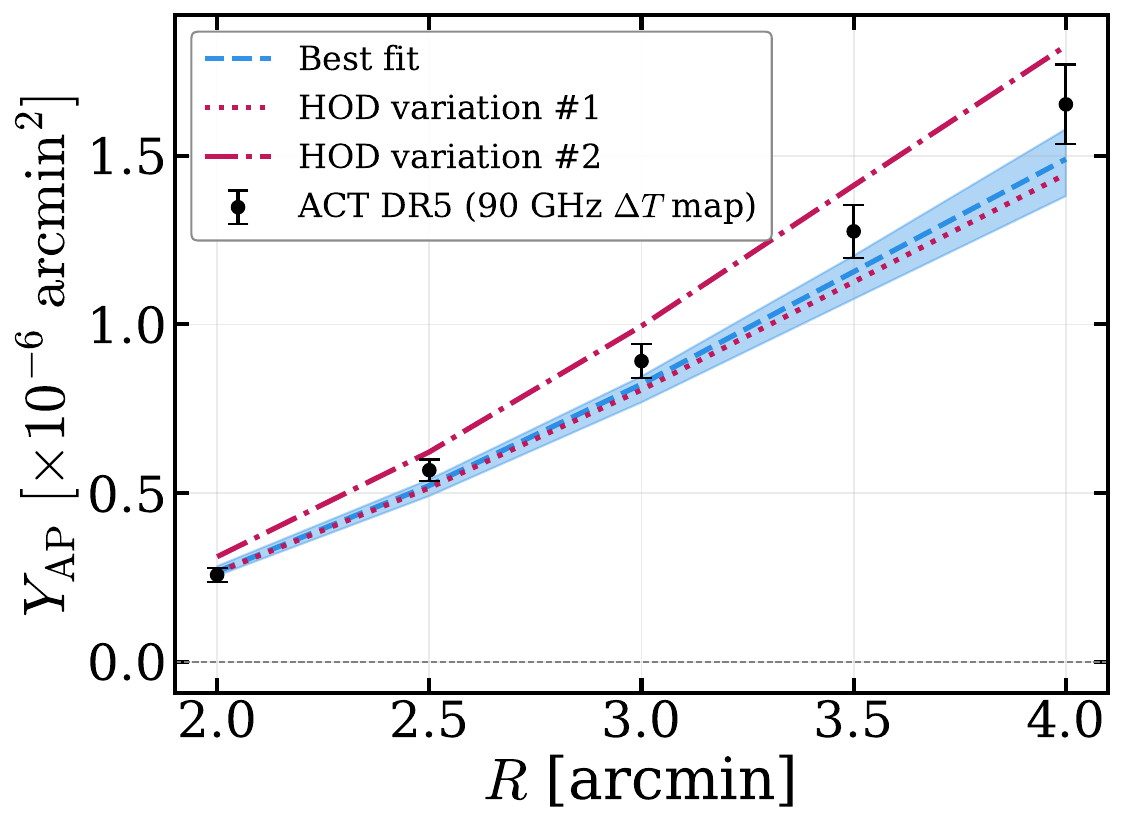} 
\caption{Stacked profile predicted at the best-fitting gNFW parameters for the fiducial HOD and two alternative HODs. HOD variation \#1 (dotted marker) corresponds to: $\log M_1 = 13.68$, $\alpha = 0.92$, $\log M_{\rm cut} = 12.72$, $\kappa = 0.39$ whilst HOD variation \#2 (dash-dot marker) uses $\log M_1 = 13.85$, $\alpha = 1.21$, $\log M_{\rm cut} = 12.65$, $\kappa = 1.45$. Both HOD variations considered here yield the same mean host halo mass as our fiducial HOD, illustrating that the tSZ signal is sensitive not only to the mean halo mass but also to the shape of the underlying halo mass distribution. This highlights the importance of allowing for greater flexibility in the HOD parametrization and incorporating complementary observables (e.g., clustering, lensing, and higher-point functions) to constrain the HOD when fitting the tSZ signal.} \label{fig:hod_variations} 
\end{figure}

\subsubsection{HOD Dependence\label{sec:hod}}

The stacked tSZ signal additionally depends on the assumed galaxy-halo connection, as it controls the high-mass tail of the galaxy population. The fiducial HOD parameters adopted in this analysis carry both statistical and modelling uncertainties, with degeneracies between $\log M_{\mathrm{cut}}$, $\log M_1$, and $\alpha$ reported in Ref. \cite{Yuan2024DESIHOD}.  Figure~\ref{fig:hod_variations} illustrates this sensitivity, showing the stacked profiles at the best-fitting gNFW parameters for two alternative HODs. These were constructed by varying $\log M_{\mathrm{cut}}$, $\log M_1$, and $\alpha$ and $\kappa$ within $\pm5\sigma$ of their fiducial values, whilst preserving the mean halo mass of the sample within $0.1 \rm dex$. The stacked profiles show variations larger than the posterior $1 \sigma$ band. Jointly varying the gNFW and HOD parameters is a necessary step toward breaking degeneracies between the gas profile and the galaxy--halo connection and obtaining robust constraints on the underlying gas physics. We leave such a joint analysis, incorporating complementary observables to constrain the HOD, to future work.







\section{Conclusions\label{sec:conclusions}}

In this work, we developed \texttt{AbacustSZ.py}, a map-level, simulation-based framework for generating tSZ maps from \textsc{Abacus} \textit{N}-body simulations, and made it publicly available. It paints gNFW pressure profiles onto light-cone halos, truncates them at a mass-dependent closure radius, assigns gas in overlapping regions to a single halo, and preserves the projected morphology of each halo using the particle maps. \texttt{AbacustSZ.py} enables fast forward modelling of mock tSZ observables, providing an intermediate approach between expensive hydrodynamical simulations and simplified analytical halo-based models.

Using tSZ maps generated with \texttt{AbacustSZ.py}, we forward-modelled the recent ACT DR5 f090 $\times$ DESI LRG stacked tSZ measurement. The simulated maps were beam-smoothed, populated with mock LRGs using \texttt{AbacusHOD}, and analysed with the same ring-ring aperture filter as the measurement. By constructing a Gaussian-process emulator and performing Bayesian inference over the gNFW pressure-profile parameter space, we obtained constraints on the gas pressure distribution surrounding DESI LRG host halos, finding a strong preference for shallow pressure profiles.

Owing to its map-level nature, \texttt{AbacustSZ.py} can also be applied to higher-order tSZ statistics, and its retention of the tSZ morphology makes it a promising tool for studying the connection between hot gas and the cosmic web through oriented tSZ stacking \cite{Lokken2022, Lokken2025}. The pressure profiles constrained here can further be combined with kSZ measurements of the same galaxies, which indicate feedback-driven redistribution of baryons to large radii \cite{hadzhiyska2025evidencelargebaryonicfeedback}. Such a joint analysis constrains the electron temperature profile, $T_e \propto P_e/n_e$, and thereby the thermal impact of feedback, helping to discriminate between feedback prescriptions \cite{Amodeo_2021}. Finally, since \texttt{AbacustSZ.py} uses the same simulations as \texttt{AbacusHOD}, the HOD and gas-pressure parameters can be varied jointly for future studies, including complementary data from lensing, clustering and/or higher-point functions, as motivated by the HOD test of Sec.~\ref{sec:hod}.



\section*{Code Availability}

\texttt{AbacustSZ.py}, together with the forward-modelling and inference code used in this work, is publicly available at \url{https://github.com/alicja-brzezowska/AbacustSZ.py} under a GLP-3.0 License. 

\section*{Acknowledgments}

We thank J.~Colin Hill and Simone Ferraro for valuable discussions. We thank William Coulton and Lehman Garrison for providing the BACKLIGHT simulation. All computations were performed using the Cambridge Service for Data Driven Discovery (CSD3), operated by the University of Cambridge Research Computing Service. Code development and debugging were aided by Anthropic's Claude Code.

\bibliography{refs}

\newpage

\appendix

\section{Elliptical halo footprints\label{app:halo_shapes}}

The halo light-cone catalogues provide, for the largest L2 subhalo of each halo, the square roots of the eigenvalues of its reduced (unit-vector-weighted) inertia tensor, $\sigma_1\geq\sigma_2\geq\sigma_3$, alongside the corresponding eigenvectors $(\mathbf{e}_a,\mathbf{e}_b,\mathbf{e}_c)$ \cite{10.1093/mnras/stab2980,10.1093/mnras/stab3066}. We use these to define the axis ratios $q=b/a=\sigma_2/\sigma_1$ and $s=c/a=\sigma_3/\sigma_1$, and set the semi-axes to
\begin{equation}
(a,b,c)=\frac{R_{200c}}{(qs)^{1/3}}\,(1,q,s),
\end{equation}
such that $abc=R_{200c}^3$, preserving the volume of the spherical halo. All three semi-axes are then multiplied by $R_c/R_{200c}$ (Sec.~\ref{sec:signal_morph}), and converted to angles at the comoving distance of the halo. Writing the halo ellipsoid in quadratic form, we obtain
\begin{equation}
\mathbf{x}^{\rm T}\mathsf{Q}\,\mathbf{x}=1,\qquad
\mathsf{Q}=\frac{\mathbf{e}_a\mathbf{e}_a^{\rm T}}{a^2}+\frac{\mathbf{e}_b\mathbf{e}_b^{\rm T}}{b^2}+\frac{\mathbf{e}_c\mathbf{e}_c^{\rm T}}{c^2},
\end{equation}
where $\mathbf{x}$ is the position relative to the halo centre.

For a halo observed along the direction $\hat{\mathbf{n}}$, let $(\mathbf{t}_1,\mathbf{t}_2)$ form an orthonormal basis of the plane perpendicular to $\hat{\mathbf{n}}$. The painted footprint is then given by the ellipse in which this plane, taken through the halo centre, intersects the ellipsoid, and is described by the $2\times2$ matrix
\begin{equation}
S_{ij}=\mathbf{t}_i^{\rm T}\mathsf{Q}\,\mathbf{t}_j,\qquad i,j\in\{1,2\}.
\end{equation}
A pixel in direction $\hat{\mathbf{p}}$ has gnomonic tangent-plane coordinates $u=\hat{\mathbf{p}}\cdot\mathbf{t}_1/\hat{\mathbf{p}}\cdot\hat{\mathbf{n}}$ and $v=\hat{\mathbf{p}}\cdot\mathbf{t}_2/\hat{\mathbf{p}}\cdot\hat{\mathbf{n}}$, and is painted if it satisfies
\begin{equation}
S_{11}u^2+2S_{12}uv+S_{22}v^2\leq1.
\end{equation}
The eigenvectors of $\mathsf{S}$ determine the orientation of the ellipse on the sky, whilst the inverse square roots of its eigenvalues determine the corresponding projected semi-axis lengths. As for circular footprints, each pixel receives the spherical profile at its angular separation from the halo centre, redistributed as in Eq.~(\ref{eq:redistribution}). The ellipse only selects which pixels are painted. Both elliptical and circular footprints are available in \texttt{AbacustSZ.py}.

\section{Semi-analytic halo model of the stacked profile\label{app:halo_model}}

As a cross-check of the simulation-based forward model (Sec.~\ref{sec:forward_model}), we additionally predict the same stacked profile using a halo model, in which the one-halo and two-halo contributions are computed analytically. We follow the approach of Amodeo et al.~\cite{Amodeo_2021} and their public code \texttt{Mop-c-GT}, whose two-halo term is based on the halo model of Vikram et al.~\cite{Vikram2017} in the formalism of Cooray \& Sheth~\cite{CooraySheth2002}, and which reproduces the compensated aperture photometry of Ref.~\cite{Schaan_2021}. The two models share the gNFW pressure profile with its mass and redshift scalings (Sec.~\ref{sec:theory}), the truncation at the closure radius, the LRG host halos, the beam and the ring-ring filter.

\subsection{Model}

The predicted stacked profile at aperture $\theta_d$ is
\begin{equation}
\label{eq:sa_model}
Y(\theta_d)=Y_{\rm 1h}(\theta_d)+A_{\rm 2h}\,Y_{\rm 2h}(\theta_d),
\end{equation}
where $Y_{\rm 1h}$ is the signal of the LRG host halos, $Y_{\rm 2h}$ the signal of correlated neighbouring halos, and $A_{\rm 2h}$ a free amplitude of the two-halo term, as in Ref.~\cite{Amodeo_2021}. 

\subsection{One-halo term}

For a host halo of mass $M\equiv M_{200c}$ at redshift $z$, the Compton-$y$ profile at angular separation $\theta$ follows from Eq.~(\ref{eq:compton_y}),
\begin{equation}
\label{eq:sa_1h}
y_{\rm 1h}(\theta)=\frac{2\sigma_T}{m_e c^2}\int_0^{l_{\rm max}} dl\; P_e\!\left(\sqrt{l^2+b^2}\right),
\end{equation}
with physical impact parameter $b=d_A(z)\,\theta$ and $l_{\rm max}=(R_t^2-b^2)^{1/2}$.
The pressure is thus truncated in a sphere of radius $R_t=R_c(M)$ (Sec.~\ref{sec:profile_generation}).

We subsequently place satellite LRGs on the projected NFW profile \cite{Navarro1997UniversalProfile} of the host, truncated at $R_{200c}$, with the concentration--mass relation of Diemer \& Joyce~\cite{DiemerJoyce2019}. The mean profile around a satellite is then given by the host profile convolved with this offset distribution,
\begin{equation}
\label{eq:sa_sat}
y_{\rm sat}(\boldsymbol{\theta})=\int d^2\theta'\;p(\boldsymbol{\theta}')\;y_{\rm 1h}(|\boldsymbol{\theta}-\boldsymbol{\theta}'|),
\end{equation}
where $p$ is the projected NFW density of the host, normalized to unity.

Weighting each host by its mean numbers of central and satellite LRGs, $\langle N_c\rangle$ and $\langle N_s\rangle$, the mean one-halo profile around an LRG is
\begin{equation}
\label{eq:sa_avg}
\begin{split}
\bar{y}_{\rm 1h}(\theta)={}&\frac{1}{\bar{n}_g}\int dM\,n(M)\Big[\langle N_c\rangle\,y_{\rm 1h}(\theta\,|\,M)\\
&+\langle N_s\rangle\,y_{\rm sat}(\theta\,|\,M)\Big],
\end{split}
\end{equation}
with $n(M)\equiv dn/dM$ and $\bar{n}_g=\int dM\,n(M)\,\langle N_g\rangle$, where $\langle N_g\rangle=\langle N_c\rangle+\langle N_s\rangle$.

\subsection{Two-halo term}

The two-halo term describes the pressure of other halos, correlated with the LRG host through large-scale structure \cite{Vikram2017,Hill_2018}. Its contribution to the galaxy--pressure cross-power spectrum is given by
\begin{equation}
\label{eq:sa_p2h}
\begin{split}
P_{gP}(k)&=b_{\rm LRG}\,P_{\rm lin}(k)\int dM\,n(M)\,b(M)\,u_P(k|M),\\
b_{\rm LRG}&=\frac{1}{\bar{n}_g}\int dM\,n(M)\,b(M)\,\langle N_g\rangle,
\end{split}
\end{equation}
with all quantities evaluated at the LRG redshift $z$. Here $n(M)$ and $b(M)$ are the halo mass function and linear bias of Ref.~\cite{Tinker2010}, and $P_{\rm lin}$ is the linear matter power spectrum with the transfer function of Ref.~\cite{Bardeen1986}, all evaluated for the simulation cosmology (Table~\ref{tab:cosmology_parameters}). The function
\begin{equation}
\label{eq:sa_uP}
u_P(k\,|\,M)=\int_0^{R_t}dr\;4\pi r^2\,\frac{\sin kr}{kr}\,P_{\rm th}(r\,|\,M)
\end{equation}
is the Fourier transform of the truncated gNFW profile. The mass integral runs from $10^{11.5}\,M_\odot$ to the cluster-mask threshold of Sec.~\ref{sec:forward_model}. The galaxy--pressure correlation function,
\begin{equation}
\label{eq:sa_xi}
\xi_{gP}(r)=\int_0^\infty\frac{dk\,k^2}{2\pi^2}\,\frac{\sin kr}{kr}\,P_{gP}(k),
\end{equation}
is subsequently projected along the line of sight as in Eq.~(\ref{eq:sa_1h}), with $P_e$ replaced by $(P_e/P_{\rm th})\,\xi_{gP}$ and the integral extended to $|l|=600$\,Mpc, giving $y_{\rm 2h}(\theta\,|\,z)$.

\subsection{Forward model}

We take the host halos of the LRGs from the \texttt{AbacusHOD} mocks of Sec.~\ref{sec:forward_model}, which sample $n(M)\langle N_c\rangle$ and $n(M)\langle N_s\rangle$ for the simulation halos. We convert their host masses to $M_{200c}$ as in Sec.~\ref{sec:abacustsz}, assign each LRG the redshift of its halo-catalogue snapshot. The LRGs are then binned in 150 bins of $\log_{10}M$ and 15 bins of redshift over $0.49<z<0.79$. The galaxy-weighted integrals, Eq.~(\ref{eq:sa_avg}) and $b_{\rm LRG}$, then become sums over the numbers of central and satellite LRGs in each bin, which also average both terms over the LRG redshift distribution.
Each profile is convolved with the Gaussian f090 beam (FWHM 2.1\,arcmin) and passed through the ring-ring filter of Eq.~(\ref{eq:ringring}), giving $Y(\theta_d)$ in ${\rm arcmin}^2$, the same quantity as the simulation stack.

All ingredients which do not depend on the gNFW parameters, namely the quadrature weights, satellite convolutions, beam and filter matrices and the two-halo kernel, are computed only once. Since $Y_{\rm 1h}$ and $Y_{\rm 2h}$ are linear in the pressure profile, a single evaluation of Eq.~(\ref{eq:sa_model}) then takes approximately 0.1\,s. We evaluate the model directly at every likelihood call, without requiring an emulator, using the likelihood of Eq.~(\ref{eq:likelihood}) with $\mathsf{C}=\mathsf{C}_{\rm data}$. The free parameters are $\alpha$, $\beta_0$, $\log_{10}P_0$ and $A_{\rm 2h}$, with the priors of Table~\ref{tab:gnfw_ranges} and a uniform prior $0\leq A_{\rm 2h}\leq12$.
Since the inner slope $\gamma$ is unconstrained in this model, likely because of its increased degeneracy caused by the extra degree of freedom of  $A_{\rm 2h}$, it is fixed at its fiducial value of $-0.3$. The posterior is sampled using \texttt{dynesty} with 1500 live points, multi-ellipsoid bounds and random-slice sampling with 10 slices, and a termination criterion of $\Delta\ln Z<0.1$.

\begin{figure*}[t]
    \centering
    \includegraphics[width=1\linewidth]{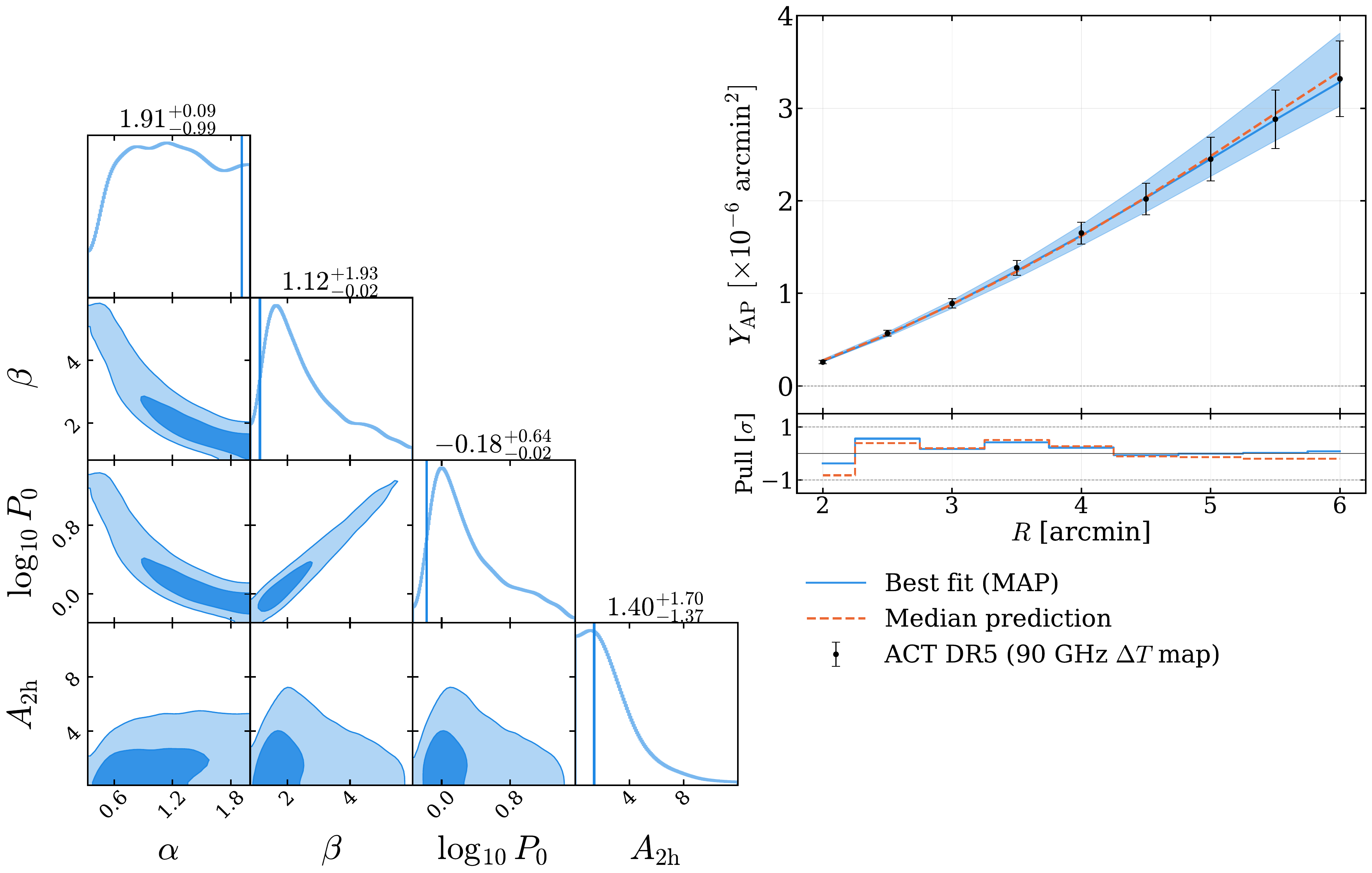}
    \caption{Posterior constraints on the gNFW pressure-profile parameters and the corresponding best-fit stacked tSZ profile obtained with the semi-analytic halo model of this appendix, which follows Amodeo et al.~\cite{Amodeo_2021}. The corner plot shows the marginalized posterior distributions of the transition sharpness $\alpha$, fall-off exponent $\beta_0$, and normalization $\log_{10}P_0$ inferred from the stacked tSZ measurements. The profile panel shows the maximum-a-posteriori (MAP) prediction (red), together with the corresponding $68\%$ credible interval (blue shaded region) and the normalized residuals, $(\mathrm{data}-\mathrm{model})/\sigma_{\mathrm{data}}$.}
    \label{fig:semianalytical}
\end{figure*}

\subsection{Results}

Figure~\ref{fig:semianalytical} shows the posterior and the best-fitting profile. The model gives $\chi^2 = 3.37$ and $N_{\rm eff}=3.9$, and hence ${\rm PTE}=0.65$ for 5.1 effective degrees of freedom, with $N_{\rm eff}$ estimated as in Sec.~\ref{sec:inference} for $\mathsf{C}=\mathsf{C}_{\rm data}$. It fits the data with smaller residuals than the simulation-based model, but has more freedom: $A_{\rm 2h}$ scales the two-halo term independently of the gNFW profile, and an enhanced two-halo amplitude of $A_{2h} = 1.4$ is preffered in this model, dominating the fit at larger apertures. Despite this difference, the two models recover broadly consistent pressure-profile constraints, which serves as a consistency check of the \texttt{AbacustSZ.py} methodology.


.



\end{document}